\documentclass[onecolumn,prd,aps,floats,floatfix,nofootinbib,11pt]{revtex4}
\usepackage{verbatim,graphicx,amssymb,amsbsy,bm,amsmath,rotating,epsfig}
\usepackage{psfrag}
\usepackage{hyperref}
\usepackage{fontenc}
\usepackage[latin1]{inputenc}
\usepackage{pstricks,pst-node,pst-text,pst-3d}
\usepackage{textcomp}
\newcommand{\be}{\begin{equation}}
\newcommand{\ee}{\end{equation}}
\newcommand{\bea}{\begin{eqnarray}}
\newcommand{\eea}{\end{eqnarray}}

\newcommand{\br}{\ensuremath{\mathbf{r}}}
\newcommand{\bv}{\ensuremath{\mathbf{v}}}
\newcommand{\bw}{\ensuremath{\mathbf{w}}}

\begin{document}
\title{Galaxy phase-space density data preclude Bose-Einstein condensate be the total Dark Matter}
\author{\bf H\'ector J. de Vega $^{(+)}$} 
\author{\bf Norma G. Sanchez $^{(a)}$}
\email{Norma.Sanchez@obspm.fr} 
\affiliation{$^{(+)}$ CNRS LPTHE, Sorbonne Universit\'e, Universit\'e Pierre et Marie Curie UPMC, Paris, Cedex 05, France. \\
$^{(a)}$ CNRS  LERMA PSL-Observatoire de Paris, Sorbonne Universit\'e  \\ 
and The Chalonge - de Vega  International School Center, Paris, France.}

\date{\today}
\begin{abstract}
 Ultralight scalars with typical mass of the order $ m \sim 10^{-22} $ eV and light scalars forming a Bose-Einstein condensate (BEC)
exhibit a Jeans length in the kpc scale and were therefore proposed as dark matter (DM) candidates. 
Our treatment here is generic, independent of the particle physics model and applies to all DM BEC,
in  both : in or out of equilibrium situations.
Two observed quantities crucially constrain DM in an inescapable way: the average DM density $ \rho_{DM} $ and
the phase-space density $ Q $. The observed values of $ \rho_{DM} $ and $ Q $ in galaxies today,
constrain both the possibility to form a BEC and the DM mass $ m $. These two constraints robustly 
exclude axion DM that decouples after inflation. Moreover, the value $ m \sim 10^{-22} $ eV 
can only be obtained with a number of ultrarelativistic degrees of freedom at
decoupling in the trillions which is impossible for decoupling in the radiation dominated era.
In addition, we find for the axion vacuum misalignment scenario that axions are produced
strongly out of thermal equilibrium and that the axion mass in such scenario
turns to be {\bf 17 orders of magnitude} too large to reproduce the observed galactic structures.
Moreover, we also consider inhomogenous gravitationally bounded BEC's supported by the bosonic quantum pressure 
independently of any particular particle physics scenario. For a typical size $ R \sim $ kpc and compact object 
masses $ M  \sim 10^7 \; M_\odot $they remarkably lead to the same particle mass  $ m \sim 10^{-22} $ eV  as the BEC 
free-streaming length. However, the phase-space density for the gravitationally bounded BEC's turns to be more than 
{\bf sixty orders of magnitude} smaller than the galaxy observed values. We conclude that the BEC cannot 
be the total DM. The axion can be candidate to be only part of the DM of the universe.
Besides, an axion in the mili-eV scale may be a relevant source of dark energy through 
the zero point cosmological quantum fluctuations.

\bigskip

(+) passed away https://chalonge-devega.fr/HdeV.html\\
(a): https://chalonge-devega.fr/sanchez/
\end{abstract}
\pacs{95.35.+d, 98.80.-k,14.80.Va}
\keywords{Dark Matter, Axions}

\maketitle
\tableofcontents

\section{Introduction}

Deciphering the nature of dark matter (DM) is nowadays one of the most active domains in
astrophysics, cosmology and particle physics. Cold dark matter (CDM) particles heavier than a GeV   succeed to reproduce the observations for large
scales beyond the Mpc. DM particles with mass $ m $ below the eV (HDM-hot DM) are ruled
out because their too large Jeans lengths exclude the
formation of the observed galaxies. There is a way out for scalar particles
if they form Bose-Einstein condensates (BEC) where the Jeans length
can be estimated as \cite{hubagru,sik}
\be\label{jeans}
\lambda_J \sim 4 \; \sqrt{\frac{10^{-22} \; {\rm eV}}{m}} \; {\rm kpc} \simeq
1.2 \times 10^{17} \; \sqrt{\frac{10^{-22} \; {\rm eV}}{m}} \; {\rm km} \; .
\ee
 In BEC dark matter, in order to reproduce the observed galactic structures, one should have typically \cite{hubagru,sik} : 
\be\label{22}
m \sim 10^{-22}  \; {\rm eV} \;.
\ee

In the linear approximation, the mass value is constrained up to $10^{-26}$ eV approximately; the value $10^{-22}$ eV arises from a non-perturbative analysis of small scales.

The same requirement but for non-BEC dark matter gives $ m $ in the keV scale, 
that is warm dark matter (WDM) \cite{ddvs},\cite{highm},\cite{highp},\cite{cosmoprd},\cite{evolwdm},\cite{moreno},\cite{boya},\cite{newas},\cite{astro},\cite{urc},\cite{eqesta},\cite{eddi},\cite{menci2016},\cite{whitep},\cite{universe2021}. CDM and WDM yield identical results 
for large scales beyond the Mpc; WDM provides too the correct medium, galactic  and small scales in agreement with observations Refs \cite {ddvs} to \cite{universe2021} and references therein.

\medskip

BEC of alkali atoms, BEC of molecules and BEC of magnons have been observed experimentally in
the laboratory \cite{leg}.

\medskip

The galactic phase space density is an important physical  quantity and its analysis is crucial to constrain the nature of Dark Matter from the by now robust observational data for it, as described in Section III and IV here below and references therein.

In this paper, we study the density in physical space $\rho$ and the density in phase-space $Q$ in order to constrain for the first time with these two observables the Bose-Einstein condensates (BECs) as Dark Matter candidates, mainly the mass of generic light scalars forming a DM BEC and other relevant properties of such BECs. 
 
\medskip
 
 Axion cosmological scenarii and BECs as Dark matter have a wide literature from many years see for example \cite{kim},\cite{marsh},\cite{balleste},\cite{borsanyi},\cite{sakharov},\cite{khlopov},
 and is not our aim here to review all them,  this is not a review paper, and our aim here is to provide constraints never considered before for DM BECs. We stress that previous DM BEC literature have not introduced the modern DM galactic phase density and the constraints imposed by the real galaxy data for the  DM phase density, and none of previous DM BEC papers relate to the aspects of the BEC DM constraints we are treating for the first time in BEC DM here.

Building particle physics models of axion scenarii although interesting in its own right, is not the aim of this paper. The DM BEC phase space density constraints and the galaxy data for them are not treated  in the previous DM BEC literature.

\medskip

Moreover, the phase space density and its galactic data to constraint DM is a treatement rather generic and universal, independent of the details of such models. This apply to any kind of DM and is manifest in the astrophysical and dark matter galactic literature. 

\medskip

In this paper we consider and constraint with galaxy data in different and consistent ways different observables and physical magnitudes as:
\begin{itemize}
\item{the density in real space,} 
\item{the phase space density,} 
\item{the surface density,}  
\item{the free streaming length,} 
\item{the number of effective degrees of freedom,} 
\item{the mass range of the different Ultralight mass particles in the BECs,}
\item{in different situations, in thermal and out of thermal equilibrium, for homogeneous as well as for gravitational non homogeneous BECs.}
\item{Cross-correlation and self-consistency of the obtained constraints for all the above relevant physical magnitudes make the results of this paper strongly robust and far beyond the literature in the field.}
\end{itemize} 

The {\it two} observables: the average DM density $\rho_{DM}$ {\it and} the galactic phase space density $Q$ robustly constrain in an inescapable way both: the possibility to form a BEC, e.g $(T_d/T_c)$, and the DM particle mass $ m $, ruling out BEC DM in general, and the BEC axion DM in particular. Moreover, the typical value $ m \sim 10^{-22} $ eV can only be obtained with a number of ultrarelativistic degrees of freedom at decoupling in {\bf the trillions} which is impossible for decoupling in the radiation dominated era. The situation for lighter DM particles is even worst and makes the exclusion result even stronger. 

\medskip

This paper is organized as follows: 

\medskip

In Section II we analyze the Bose Einstein Condensate as a Dark Matter candidate. Our treatment applies to any shape of the distribution function and is valid for any particle physics model. 
A detailed and updated analysis of the BEC phase space density and BEC coarse-grained phase space density is provided in Sections III and IV, both in theory and observations. 

Section III provides an updated  synthesis and clarification on the phase density in DM and a useful state-of-the art. From such analysis,  robust DM BECs constraints are derived in Section IV. 

DM BEC decoupling at Thermal Equilibrium is treated in Section V including its implications for the DM axion. DM BEC in decoupling out of Thermal Equilibrium is treated in Section VI, including its implications for the typical Jeans- Lengths and for the BEC number of ultra-relativistic degrees of freedom. 

In Section VII we discuss inhomogeneous, gravitationally bounded BE condensates of finite size. The constraints from the galactic phase space density remarkably provide similar values to the homogeneous BEC constraints found in Sections IV and V and confirm the generic character and robustness of the results. 

In Section VIII we analyze DM axions in the canonical axion vacuum misalignment scenario \cite{kt}, \cite{tres}, \cite{mt} and we constraint it with the DM phase space density and galactic data. The exclusion constraints obtained for the axion as a DM candidate confirm the universal constraints obtained in this paper for the DM BECs in general. 

Section IX summarizes our results, conclusions and remarks. We notice that  axions with masses in the meV = $10 ^{-3}$ eV range  can play an important role in astrophysics and cosmology, \cite {de}, \cite{isern}, \cite{wang}, \cite{bonnaz},  \cite{bianchi}, not for dark matter but for dark energy as we proposed and studied in ref \cite {de},  and the misalignment scenario could
produce axions with mass in such meV scale.
 
\section{The Bose Einstein Condensate (BEC) as a Dark Matter candidate}

\begin{itemize}
\item{ After decoupling, the DM distribution function freezes out and
is a function of the covariant momentum $ p $. We consider \textbf{generic} distribution functions $ f_d $ out of thermal equilibrium or thermal. The specific form of $ f_d $ in the non-thermal cases depends on the details of the interactions before decoupling.} 
\item{ Our treatment applies to \textbf{any shape} of $ f_d $ and is valid for \textbf {any} particle physics model. For convenience and without loss of generality, we choose $ f_d $ as a function of $ (p/T_d) : \;  f_d \;(p/T_d) $,
where $ T_d $ is the covariant decoupling temperature.}
\end{itemize}
In a BEC a sizeable fraction of the particles is in the zero momentum state
while the rest is on excited states. We call $ \rho_0 $ the zero-momentum comoving contribution
to the mass density. The contribution from the excited states $ \rho - \rho_0 $ follows as usual by integrating
the distribution function. When the particles became nonrelativistic, we thus have
\be\label{resto}
\rho - \rho_0 = m \; \int_0^{\infty} \frac{p^2 \; dp}{2 \; \pi^2} 
 f_d\left(\frac{p}{T_d}\right) = \frac{m}{2 \; \pi^2} \; T_d^3 \; U  \quad, 
\ee 
\be\label{restob}
U  \equiv \int_0^{\infty} y^2 \; f_d(y) \; dy \; ,
\ee
where we consider neutral scalars. The case where the particles remain ultra-relativistes (UR) is considered 
in eq.(\ref{ultrar}) below. 

The BEC density $ \rho_0 $ vanishes at the BEC covariant critical temperature $ T_c $. 
Therefore, the BEC can be present if $ T_d < T_c $ and we have from eq.(\ref{resto}) \cite{bdvs},
\be\label{rho}
\rho = \frac{m \; U}{2 \; \pi^2} \; T_c^3 \quad , \quad 
\rho_0 =\frac{m \; U}{2 \; \pi^2} \;\left(T_c^3 - T_d^3 \right) \quad , \quad T_d < T_c \; .
\ee
[$ T_c $ is defined by the above equation even in the out of thermal equilibrium case].
$ T_d $ and $ T_c $ are related to the respective effective
number of UR degrees of freedom $ g_d $ and  $ g_c $, and to the photon 
temperature today $T_{\gamma}$ by entropy conservation \cite{kt}: 
\be\label{temp}
T_c = \left(\frac2{g_c}\right)^{1/3} \; T_{\gamma}  \; , \quad 
T_d = \left(\frac2{g_d}\right)^{1/3} \; T_{\gamma} \; , \quad T_{\gamma} = 0.2348 \;  10^{-3} {\rm eV} \; .
\ee
The DM density $ \rho $ must reproduce the observed average DM in the universe $ \Omega_{DM} \; \rho_{\rm crit} $. 
Hence,

\be\label{rhoDM}
\rho_{\rm DM}\; \equiv \;\Omega_{DM} \; \rho_{\rm crit}\; = \;\frac{m \; U}{\pi^2 \; g_d} \; 
\; \left(\frac{T_c}{T_d}\right)^3 \; T_{\gamma}^3 ,
\ee
$$
\rho_{\rm DM} = 0.9259 \; 10^{-23} \; {\rm keV}^4 \; 
$$

Therefore, the DM particle mass $ m $ can be related to $( T_d/T_c )$ as

\be\label{m}
m = \pi^2 \; \frac{\rho_{DM} \; g_d }{T_{\gamma}^3 \; U} \; \left(\frac{T_d}{T_c}\right)^3 =
7.059 \; {\rm eV} \; \frac{g_d }{U} \; \left(\frac{T_d}{T_c}\right)^3 \; .
\ee

{\vskip 0.1cm}

The value of $ g_d $ depends on the detailed particle physics of the light scalar particle.
For QCD axions decoupling soon after the QCD phase transition one has $ g_d \sim 25 $.
The covariant critical temperature $ T_c $ as well as $ g_c $ are parameters that
depend on the BEC state. 
$$
\mbox {It must be  $g_c < g_d$  (and hence $ T_d < T_c $) in order to have a BEC.}
$$
{\vskip 0.3cm}

The continuous and bounded function $ f_d\left(p /T_d \right) $ stands for
the excited states of the distribution function. The total distribution function
can be written as a Dirac delta function representing the zero momentum BEC 
plus the excited states piece $ f_d\left(p /T_d \right) $ as follows

\be\label{deld}
f_d^{total}(p) = (2 \; \pi)^3 \; \frac{\rho_0}{m} \; \delta\left(\vec{p}\right) +
 f_d\left(\frac{p}{T_d}\right) = 2 \; \pi^2 \; \frac{\rho_0}{m} \; \frac{\delta(p^2)}{p^2} +
 f_d\left(\frac{p}{T_d}\right)
\ee

{\vskip 0.1cm}

The continuum Dirac delta notation is convenient for calculations but in reality the BEC is in a finite 
comoving volume $ V_c $ and the wavenumbers $ \vec{p} $ are discretized:
$$
\vec{p} = \frac{2 \; \pi}{ V_c^{1/3}} \; \vec{n} \quad , \quad \vec{n} \; \epsilon \; {\cal Z}^3  \quad , \quad 
\delta\left(\vec{p}\right) = \frac{V_c}{(2 \; \pi)^3} \; \delta_{\vec{p},\vec{0}} \; .
$$

Therefore, $ \delta (\vec{0}) = \displaystyle \frac{V_c}{(2 \; \pi)^3} $ is  finite  as  well as $ f_d^{total}(0) $:
$$
f_d^{total}(0) = \frac{\rho_0}{m} \; V_c + f_d (0) \; .
$$
The comoving square velocity $ <v^2> $ can be then expressed as

\be\label{v2}
<v^2>\; =\; \frac{<p^2>}{m^2}\; =\; \frac{1}{m^2} \; \frac{\int_0^{\infty} p^4 \;   
 f_d\left(p /T_d \right)\; dp /(2 \; \pi^2)}{(\rho_0 / m) + \int_0^{\infty} 
p^2 \; f_d\left(p /T_d\right) \; dp / (2 \; \pi^2)} 
\ee
\be\label{v2b}
<v^2>\; = \;\frac{T_d^5}{m^2 \; T_c^3} \; \frac{V}{U} \;,
\quad \quad V \equiv \int_0^{\infty} y^4 \; f_d(y) \; dy \;
\ee

and $ U$ is given by eq.(\ref{restob}). The BEC does not contribute to the integral in the numerator of $<v^2>$ in eq.(\ref{v2}) because the integral of the Dirac delta function in $f_d^{total}$  eq.(\ref{deld}) vanishes upon integration over 
$ \vec{p} $ due to the extra $ p^2 $ factor in the numerator of eq.(\ref{v2}) with respect to the denominator.

{\vskip 0.3cm}

As noticed in ref. \cite{bdvs} the Bose-Einstein distribution,
 for massless particles is of the order of the comoving volume 
$ V_c $ as discussed above.
The BEC distribution function is well defined.
This is also the case for massive particles because the chemical potential at decoupling $ \mu_d $ must be equal to the particle mass $ m $ in order to form a BEC \cite{bdvs}. 
Even the part of the distribution function that describes the particles outside the condensate is of the order of $ V_c $.

Therefore, the two pieces of the total distribution function: the BEC part and the excited states part $ f_d\left(p /T_d\right) $ are of the order of the comoving volume $ V_c $ which in this case is very small.
This is so because for the QCD phase transition $ g_d \sim 25 $ and $ (z_d + 1) \sim 1.7 \; 10^{12} $, giving for $ V_c $,
\be
V_c \;= \;\frac{V_{today}}{(z_d + 1)^3}\; \sim \;2 \times  10^{-37} \; V_{today} \; .
\ee

Taking for example $ V_{today}^{1/3} = 1 $ kpc, yields $ V_c^{1/3} \sim 18200 $ km. 

\section{The BEC Phase Space Density} 

Let us discuss now the phase space density which is defined by \cite{dalcanton}:

\be 
\displaystyle Q \equiv \frac{\rho}{<\sigma^2>^{3/2}}\; = \;\sqrt{27} \; \frac{\rho}{<v^2>^{3/2}} \; 
= \; \sqrt{27} \; \frac{m^3 \; \rho}{\big\langle \vec{P}^2_f \big\rangle^{3/2}} \; , \label{D}
\ee

where $ \sigma^2 = <v^2>/3 $ is the velocity dispersion, and $ \vec P_f $ is the physical momentum.

{\vskip 0.3cm}

 Including explicitly the BEC, using eqs.(\ref{rho}) and (\ref{v2}), the phase space density $ Q $ eq.(\ref{D}) is given by 
\be \label{defQ}
\displaystyle Q = \sqrt{27} \; m^3 \; \frac{ \left[(\rho_0 /m) + \int_0^{\infty} 
p^2  \; f_d \left(p /T_d\right) \; dp / (2 \; \pi^2) \right]^{5/2}}{\left[\int_0^{\infty} p^4 \;f_d\left(p / T_d)\right) \; dp / (2 \; \pi^2) \right]^{3/2}}
\ee
 $$
 Q = \frac{3 \; \sqrt3}{2 \; \pi^2} \; m^4 \; \frac {U^{5/2}}{V^{3/2}} \;\left(\frac{T_c}{T_d}\right)^{\! {15/2}} \; 
$$
\begin{itemize}
\item{In the absence of self-gravity $ Q $  is Liouville invariant 
because both $ \rho $ and $ \big\langle \vec{P}^2_f
\big\rangle^{3/2} $ redshift as $ (z+1)^3 $.} 
\item{Because the distribution function is frozen and is a
solution of the collisionless Boltzmann (Liouville) equation,  it is clear that $ Q $ is a \emph{constant}, namely a Liouville invariant,  in the absence of self-gravity \cite{bdvs}.}
\item{The value of $Q$ given by eq.(\ref{defQ}) is valid after decoupling and before structure formation when $Q$  is invariant under the universe expansion.}
\end{itemize}

\section{ The BEC Coarse-Grained Phase Density Constraint} 

The expression of $ Q $ eq.(\ref{D}) provides a good approximation to \textit{the coarse-grained phase-space density}.
Given the central role played by the  phase-space density in the galaxy DM context, we derive here the connection between the coarse-grained phase-space distribution function $ F(\br,\bv) $ and $ Q(\br) $.

The matter density express in terms of $ F(\br,\bv) $ as
\be \label{psd}
\rho(\br) = \int d^3 v \;  F(\br,\bv)\; =\; <v^2>^{3/2} \; \int d^3 w \;  F(\br,\bw) \; ,
\ee
where we change the integration variable to $ \bw \equiv \bv/\sqrt{<v^2>} $. 

We find from eq.(\ref{psd}) the inequality
$$
 \frac{\rho(\br)}{<v^2>^{3/2}} = \int d^3 w \;  F(\br,\bw) \;\; \geq \;\int_{|\bw|\leq 1} d^3 w \;  F(\br,\bw)
$$
where we used the positiveness of the distribution function $ F(\br,\bw) \geq 0 $.

Now we can apply the first mean value theorem \cite{gra} to the last integral over $ \bw $,
\be \label{Qpsd}
Q(\br) \equiv \frac{\rho(\br)}{<v^2>^{3/2}}\; \;\geq  \int_{|\bw|\leq1} d^3 w \;  F(\br,\bw_1) = \frac{4 \; \pi}3 
\;  F(\br,\bw_1) \; ,
\ee

where $ \bw_1 $ is a point inside the unit sphere. 
\begin{itemize}
\item{We find that  $ Q $ approximates  the coarse-grained phase-space distribution function  by excess by a factor of order one.
Previous estimations \cite{shao}  of $ Q $ yielded values similar to the rigorous derivation presented here.} 
\item{Therefore, eq.(\ref{D}) provides an appropriate coarse-grained phase-space density $ Q $. 
The  phase-space density expressions eq.(\ref{D}) are always valid:
in the primordial universe where it is constant as well as afterwards in the 
presence of self-gravity when structure formation occurs.} 
\end{itemize}
Tremaine and Gunn \cite{TG} argued that the value of the coarse grained phase space density is \textit{always smaller} than, or equal
to,  the maximum  value of the (fine grained) microscopic phase space density, which is the distribution function. 

Such argument relies on the theorem that states that the coarse grained phase space density 
can only diminish by collisionless phase mixing or violent relaxation by gravitational dynamics \cite{theo}. 

A similar argument was presented by Dalcanton and Hogan \cite{dalcanton}, and confirmed by numerical studies. 

In ref \cite{mnrasdvs} $Q$ was implemented in a model independent manner to evaluate the DM particle mass scale in non-BEC DM.

The phase-space density $ Q $ is a \emph{constant} in the absence of self-gravity, and  
$ Q $ can only {\bf decrease} by collisionless phase mixing or self-gravity dynamics  \cite{dalcanton}, \cite{TG}, \cite{theo}, \cite{madsen}. For these reasons,  $ Q^{-1} $ behaves as an entropy that can only increase or stay constant during the universe expansion.

Therefore, necessarily :
\be Q_{\rm today} \leq Q \; ,   \quad    {\rm where}  \quad  Q_{\rm today} \equiv \frac{Q}{Z} \; \quad
\ee

{\vskip 0.1cm}

being $ Z \geq 1 $ a numerical constant, namely {\it the decreasing factor} $Z$ introduced in Ref \cite{mnrasdvs}. The value of $ Q_{\rm today} $ can be computed with galaxy data today for $ \rho $ and $ <\sigma^2> $, namely 
\be 
\displaystyle Q_{\rm today} =  \frac{\rho_{\rm today} }{<\sigma_{\rm today}^2>^{3/2}} 
\ee 
    
Normally, $ \rho  $ and $ \sigma^2 $ are averaged over the galaxy core. 

$ Q_{\rm today}  $ has been well measured by different galaxy observations and it is galaxy dependent.
$ Q_{\rm today} $ is the largest for ultracompact dwarf galaxies and the smallest for large
and dilute spiral galaxies \cite{datos}, \cite{datos2}, \cite{datos3}. From the compilation of different and well established sets of galaxy data, eg in Table 1 of ref. \cite{astro} we have
\be\label{Qdato}
5 \times 10^{-6}\; < \;\left(\frac{Q_{\rm today}}{\rm keV^4}\right)^{\! {2/3}} \;< 1.4 \; .
\ee

Now, from eqs.(\ref{m}) and (\ref{defQ}) we express $m$ and $(T_d/T_c)$ in terms of $Q , \; U$ and $ V $ with the result:

\bea \label{mTdTc}
&& m \;= \; \frac{2^{2/3}}{3 \; \pi^2} \; V \;  Q^{2/3} \; 
\frac{T_{\gamma}^5}{(\rho_{DM} \; g_d)^{5/3}}\; = \;44.62 \; {\rm keV} \; \left(\frac{Q}{\rm keV^4}\right)^{2/3}  
\left(\frac{25}{g_d}\right)^{\! 5/3} V \; , \cr \cr
&& \left(\frac{T_d}{T_c}\right)^3 = \;\frac{2^{2/3}}{3 \; \pi^4} U V \;
\frac{T_{\gamma}^8}{(\rho_{DM} \; g_d)^{8/3}}\;  Q^{2/3}\; = \; 
248.43 \; \left(\frac{Q}{\rm keV^4}\right)^{\! {2/3}} U V
\left(\frac{25}{g_d}\right)^{\! {8/3}} .
\eea

{\vskip 0.2cm}

$ g_d \sim 25 $ corresponds to DM decoupling just after the QCD phase transition as it is the case for axions. 

\section{Decoupling at Thermal Equilibrium} 

For decoupling at thermal equilibrium (TE), from eqs.(\ref{resto}) and (\ref{v2}) we have 
$$
U^{TE} \;= \;2 \, \zeta(3) \;= \;2.404114 , \quad V^{TE} \;= \;24 \, \zeta(5)\;= \;24.88627 
$$
From these values, eq.(\ref{mTdTc}) yields

\be\label{msolu}
\frac{T_d}{T_c}\; = \;24.58709 \; \left(\frac{Q}{\rm keV^4}\right)^{\! {2/9}} \; 
\left(\frac{25}{g_d}\right)^{\! {8/9}}, \;
\ee
$$
 m =  1.110 \; {\rm MeV} \; \left(\frac{Q}{\rm keV^4}\right)^{\! {2/3}} \; 
\left(\frac{25}{g_d}\right)^{\! {5/3}}: \; {\rm TE} \; .
$$

{\vskip 0.1cm}

Therefore, from eqs.(\ref{msolu}) and (\ref{Qdato}) the condition $ Q \geq Q_{\rm today} $ implies 

\be\label{conBE}
\frac{T_d}{T_c} \geq 27.5 \; \left(\frac{25}{g_d}\right)^{\! {8/9}}
\;, \quad m \geq 1.55  \; {\rm MeV} \; \left(\frac{25}{g_d}\right)^{\! {5/3}} \quad : \;{\rm TE} \; .
\ee

{\vskip 0.1cm}

We see that for $ g_d \sim 25 $, is always $ T_d > T_c $ and hence \textbf{no BEC forms for TE decoupling}.\\
In order to form a BEC is necessary that the decoupling temperature $ T_d $ be below the critical temperature $ T_c $. The condition 
$$ T_d \leq T_c $$ yields from eq.(\ref{conBE})
\be\label{1020}
g_d \geq 1040 \quad : \quad {\rm BEC, ~ TE} \; .
\ee
\begin{itemize}
\item{This requires particle models possesing a {\bf huge number} of particle states and
where DM decouples presumably in the Grand Unification scale (GUT) where the number
of ultrarelativistic degrees of freedom is in the hundreds (well above the electroweak scale).}
\item{Recall that at the TeV scale in the standard model of particle
physics, $ g_d \sim 100 $ \cite{kt}. In addition, for $ g_d = 1040 $  eq.(\ref{conBE}) yields
\be\label{mTE}
m > 3.10 \; {\rm keV} \quad , \quad g_d = 1040 \;:  \quad {\rm BEC, ~ TE} \; .
\ee
This particle mass value is much larger than the DM particle mass appropriate
for BEC DM  eq.(\ref{22}) $ m \sim  10^{-22} $ eV.       
\textbf{This is a huge difference} of orders of magnitude.}
\subsection{Implications for the Axions}
\item{\textbf{QCD axions} can decouple well before the QCD phase transition, at temperatures $ T_d \sim 10^{11} $ GeV.

For $ T_d \sim 10^{11} $ GeV we can have $ g_d $ in the
hundreds and from eq.(\ref{conBE}) the axion mass $ m $ 
turns to be \textbf{in the keV scale}, a huge difference of orders of magnitude
above the mass range values for the axion mass given by the present experimental limits \cite{sik}:
$$ 6 \times 10^{-6} \; {\rm eV} \;< \; m_a \;< \;2 \times 10^{-3}  \; {\rm eV} ,$$  and even a more huge difference with respect to the typically ultra-light BEC mass value eq.(\ref{22})  $ m \sim  10^{-22} $ eV.} 
\end{itemize}
Besides, it must be noticed that an isolated system which is not integrable 
can thermalize because is an ergodic system both classically and quantum mechanically \cite{gall}.
Namely, the particle trajectories explore
ergodically the constant energy manifold in phase-space, covering it uniformly
according to precisely the microcanonical measure and yielding to a thermal situation \cite{gall}.
This is the case for \textbf{axions} and more generally for \textbf{QCD systems}. Also, it is generally
the case of self-gravitating DM particles in galaxies \cite{mnras2014} \cite{eddi}, \cite{bhwdm}.

{\vskip 0.2cm}

In order to determine how the thermalization happens, as well as the thermalization time scale and further physical features, specific calculations must be performed in the corresponding model,
see refs. \cite{sik}, \cite{DdV}, \cite{mitk}, \cite{aabe}, \cite {beraxi}, \cite{sdav}.
The methods used in these calculations are appropriately chosen for the system
considered: Boltzmann equations, classical
evolution equations, Schwinger-Keldysh approach for quantum fields, expansions
in 1/N in field theory and others methods.

\section{Decoupling out of Thermal Equilibrium}

For {\bf decoupling out of TE} we recall that
typically, thermalization is reached by the mixing of the particle modes
and the scattering between particles which redistribute the particles in phase
space as following:

The higher momentum modes are populated
by a \emph{cascade} whose wave front moves towards the ultraviolet region
akin to a direct cascade in turbulence, leaving in its wake a state
of nearly local TE but with a temperature lower than that of equilibrium \citep{DdV}. 

{\vskip 0.1cm}

Hence, when the dark matter particles at decoupling are 
not at thermodynamical equilibrium, their momentum distribution is expected to be peaked at smaller momenta than in the TE case
because the ultraviolet cascade is not yet completed \citep{DdV}. Therefore, the
distribution function at decoupling out of TE can be written as
\be \label{fuera}
f_d^{out ~ TE} (p) = \frac{f_0}{e^{\frac{p}{\xi \; T_d}}-1} \; \theta(p^0 - p) \; ,
\ee

{\vskip 0.1cm}

where $ \xi = 1 $ at TE and $ \xi \lesssim 1 $ before
thermodynamical equilibrium is attained; $ f_0 \sim 1 $ is a normalization 
factor and $ p^0 $ cuts the spectrum in the ultraviolet region not yet reached by 
the cascade. 
\begin{itemize}
\item{The above features and the distribution function out of TE eq.(\ref{fuera}) are generic and universal, the result is unique irrespective of the different ways the massive bosons forming a BEC can be out of TE, because the formation of a BEC is a unique process requiring one universal condition $ T_d \leq T_c $.}
\item{ Eq.(\ref{fuera}) describes out of equilibrium massive scalar particles \citep{DdV}.
Out of equilibrium massless scalar particles were studied in refs. \cite{mitk}, \cite{aabe}, \cite {beraxi}.}
\end{itemize}

The distribution function eq.(\ref{fuera}) yields for $ U $ and $ V $ through
eqs.(\ref{resto}) and (\ref{v2}),
$$
U^{out ~ TE} = f_0 \; \xi^3 \; U(s) \; , \quad
V^{out ~ TE} = f_0 \; \xi^5 \; V(s) \; , \quad s \equiv \frac{p^0}{\xi \; T_d} \; , 
$$
$$
U(s) \equiv \int_0^s \frac{y^2 \; dy}{e^y - 1} \quad , \quad V(s) \equiv \int_0^s \frac{y^4 \; dy}{e^y - 1}\; .
$$
{\vskip 0.2cm}
Then, from eq.(\ref{mTdTc}) the condition $ Q \geq Q_{\rm today} $ implies for {\bf decoupling out of TE}:

\bea\label{mnequi}
&& m \;\geq \;44.62 \; {\rm keV} \; \left(\frac{Q_{\rm today}}{\rm keV^4}\right)^{2/3} \; 
\left(\frac{25}{g_d}\right)^{5/3} \; V^{out ~ TE} 
\eea
\bea\label{mnequib}
&& m \;\geq \; 1.110 \; {\rm MeV} \; \left(\frac{Q_{\rm today}}{\rm keV^4}\right)^{2/3}
\left(\frac{25}{g_d}\right)^{5/3} f_0 \; \xi^5 \; \frac{V(s)}{V(\infty)} \quad , \cr \cr
&& \left(\frac{T_d}{T_c}\right)^3 \geq 1.48635 \; 10^4 \; \left(\frac{Q_{\rm today}}{\rm keV^4}\right)^{2/3} f_0^2 \; \xi^8 \; \frac{U(s) \; V(s)}{U(\infty) \; V(\infty)} : \quad {\rm out ~ of ~ TE} \; .
\eea

Typically, out of TE we have 
$$ s = {\cal O}(1)\;, \;\;  f_0 = {\cal O}(1)\;,\;\;
U(1)/U(\infty) = 0.147  , $$
$$ V(1)/V(\infty) = 0.00658 $$
From the bound eq.(\ref{mnequi}), the limiting condition $ T_d \sim T_c $ 
for the presence of a BEC is satisfied for $ s \sim 1, \; \xi \sim 0.7, \; f_0 \sim 1 $.
As a consequence, in the BEC limiting case of {\bf decoupling out of TE}, we find:
\be\label{facneq}
T_d \sim T_c \; , \; 
f_0 \; \xi^5 \; \frac{V(s)}{V(\infty)} \sim 10^{-3}  \; , \quad 
m \geq 14 \; {\rm eV} \; \left(\frac{25}{g_d}\right)^{\frac53} : \quad {\rm BEC} \;
 {\rm out ~ of ~ TE} \; .
\ee

{\bf We conclude that:}

\begin{itemize}
\item{BEC DM decoupling at thermal equilibrium requires a particle model
with a {\bf huge number} $ g_d \geq 1040 $ of particle states 
ultrarelativistic at DM decoupling [eq.(\ref{1020})]. For $ g_d = 1040 $ the particle mass must be $ m > 3 $ keV
[eq.(\ref{mTE})], that is, {\bf twenty-five orders of magnitude larger} 
than the appropriate BEC mass value eq.(\ref{22}).}
\item{BEC DM decoupling out of thermal equilibrium requires for $ g_d \sim 25 $
a particle mass $ m $ of at least 0.03 eV [eq.(\ref{facneq})].
For $ T_d \sim 10^{11} $ GeV, $ g_d $ is in the hundreds and we obtain from eq.(\ref{facneq}) $ m $
at least {\bf twenty orders of magnitude larger} than the appropriate BEC mass value eq.(\ref{22}).}
\end{itemize}
\subsection{Implications for the BEC Jeans Lengths}
From eq.(\ref{jeans}) the Jeans lengths corresponding to the above cases eqs.(\ref{mTE}) and (\ref{facneq}) are:
$$
\lambda_J ({\rm keV}) =  3.8 \times 10^4 \; {\rm km} \quad {\rm for ~ TE} \; ,
$$
$$
\lambda_J ({\rm 10 ~ eV}) = 6.9 \times 10^6 \; {\rm km} \quad {\rm for ~ out ~ of ~ TE} \; .
$$
\begin{itemize}
\item{These BEC Jeans-length values are {\bf unrealistically small} by eleven to thirteen orders of magnitude [see eq.(\ref{jeans})]
in order to form the observed galaxy structures. Namely, DM structures of {\bf all} sizes 
above these minuscule Jeans lengths will be formed in contradiction with astronomical observations. These Jeans length values are even worse than the cold DM Jeans length which is $ \sim 3 \times 10^{12} $ km.}
\item{Therefore, the BEC particle masses compatible with the DM average density and the 
DM galaxy phase-space density constraints, namely: $$ m \; >\; 3  \;\mbox{keV \; (in   TE)}\;\;\; \mbox{and} \;\;\ m \;> \;0.03 \;  \mbox{eV \; (out of TE)}, $$ have {\bf exceedingly small} Jeans lengths, results which \textbf{strongly disfavour BEC DM}.}
\end{itemize}
\subsection{Implications for the BEC number of ultra-relativistic degrees of freedom}

It is interesting to see which value of $ g_d $ corresponds to the particle mass value 
$ m \sim  10^{-22} $ eV appropriate for galaxy structure formation. We find from eqs.(\ref{22}), (\ref{conBE}) and (\ref{facneq}) that $ g_d $ must take the values
$$
g_d \sim 2 \times 10^{11} \quad {\rm TE} \quad;
\quad g_d \sim 2 \times 10^{14} \quad {\rm out ~ of ~ TE} \; .$$
\begin{itemize}
\item{These gigantic values of $ g_d $ are totally impossible for decoupling in the radiation dominated era.
Namely, these values of degrees of freedom are
absolutely unrealistic for whatever particle
physical model one considers. Hence, there is no way to realize a tiny DM mass $ m \sim 10^{-22} $ eV.}
\end{itemize}

\textbf{In the case DM stays ultra-relativistic till today}, eq.(\ref{rho}) for the DM density becomes

\be\label{ultrar}
\rho = \frac{T_c^4}{2 \; \pi^2} \;  W \quad , \quad W \equiv \int_0^{\infty} y^3 \; f_d(y) \; dy \;
\ee

This equation is valid both in the BEC case and in the absence of a BEC.

For out of thermal equilibrium decoupling we have from eq.(\ref{fuera}) that $$ W^{~out ~ TE} \;< \;W^{TE} \;= \; \pi^4/15 $$ We thus obtain from eqs.(\ref{temp}) and (\ref{ultrar}),
\be\label{ultrarb}
g_d^{TE}\; = \;0.4443 \;> \; g_d^{~ out ~ TE}
\ee
\begin{itemize}
\item{Equation (\ref{ultrarb}) {\bf cannot} be satisfied because it must always be $ g_d \geq 2 $ due to the existence of the photon. Therefore, \textbf{scalar particles which are ultra-relativistic today cannot describe the DM}.}
\item{The treatment we presented here is {\bf independent} of the particle physics model
describing the DM particle and applies to all DM BEC. All the results found here only follow from the gravitational 
interaction of the particles, their bosonic nature and the robust DM observational constraints from the average DM density $ \rho_{DM} $ and the DM phase-space density $ Q $.}
\end{itemize}

\section{Gravitationally bounded Bose-Einstein condensates of finite size}

So far, we have here considered homogeneous Bose Einstein condensate (BEC) i.e. a BEC in all the space.
Gravitationally bounded BEC's with a finite size $ R $ can also exist. That is to say, a two phase situation in which the BEC is inside the radius $ R $ and the normal phase is outside. This gravitationally bounded BEC can be considered as the final stationary state, dynamically produced by a gravitational BEC phase transition.

{\vskip 0.1cm}

The results provided by this gravitational BEC study are robust irrespective of any particular particle physics model.
A gravitationally bounded object formed by a BEC can be obtained by equating the bosonic quantum pressure and the gravitational pressure. 
 
{\vskip 0.2cm}

The BEC quantum pressure is the flux of the quantum momentum, $ P_Q =  n \; v \; p $,  $ p $  being the minimum momentum 
from the Heinsenberg principle  $ p \sim \hbar/R $ and $ n = \rho/m $ is the number density of particles. Therefore, 
\be \label{pqbec}
P_Q= \rho \; \left(\frac{\hbar}{R \; m} \right)^2= \frac3{4 \; \pi} \; \frac{\hbar^2 \; M}{R^5 \; m^2} \; ,
\ee
where $ \rho = (3 \; M)/(4 \; \pi \; R^3) $ and $ M $ is the mass of the BEC.

{\vskip 0.1cm}

For an object of radius $ R $ and mass $ M $ the gravitational pressure is 
\be
P_G = \frac{G \; M^2}{4 \; \pi \; R^4} \; .
\ee
Thus, $ P_Q =  P_G $ implies for the BEC size $ R $,
\be \label{rbec}
R= \frac{3}{G \; M} \; \left(\frac{\hbar}{m} \right)^2 = 2.861 \; 10^{-36} \; 
\frac{M_\odot}{M} \; \left(\frac{\rm eV}{m}\right)^2 \; \hbar \; {\rm kpc} \; .
\ee
The DM particle mass becomes from eq.(\ref{rbec})
\be
m = 5.349 \; 10^{-22} \; {\rm eV} \; \sqrt{\frac{10^7 \; M_\odot}{M} \; \frac{\rm kpc}{R}} \; .
\ee
That is, for a typical compact BEC object of kpc size and mass $ M \sim 10^7 \; M_\odot $ 
we obtain the BEC DM  particle mass in agreement with eq.(\ref{22}).
This remarkable result shows the consistency of the self-gravitating BEC quantum estimate  eq.(\ref{pqbec})-(\ref{rbec})
with the BEC free-streaming length eq.(\ref{22}).

\medskip

\textbf{The phase space density} $ Q= \rho / \sigma ^3 $ can be estimated following similar lines as above, namely

\be\label{Qmaxbec}
Q= \sqrt{27} \; \rho \; \left(\frac{m \; R}{\hbar}\right)^3 = 
\frac{\sqrt{27}}{4 \; \pi} \; \left(\frac{m}{\hbar}\right)^3 \; M \; ,
\ee
with the result
\be\label{Qmaxbecb}
\frac{Q}{{\rm keV}^4} = 0.461 \; 10^{-68}  \;\left(\frac{m}{10^{-22} \; {\rm eV}}\right)^3  \; \left(\frac{M}{10^7 \; M_\odot}\right) \; .
\ee

\begin{itemize}
\item {BEC objects would correspond to compact halos ie typically $ M $ about $ 10^7 \; M_\odot$,
thus $ Q \sim  10^{-68} $ for the typical $ m \sim 10^{-22} $ eV. That is, $Q$ turns out \textbf{more than sixty orders of magnitude smaller} than  the observed values eq.(\ref{Qdato}).}
\item {Although $ m \sim 10^{-22} $ eV provides reasonable BEC free-streaming lengths [eq.(\ref{22})],
the corresponding BEC phase-space density turns to be \textbf{ridiculously small}.
 {\vskip 0.1cm}
Notice that the value eq.(\ref{Qmaxbec}) is a {\bf maximal} value for $ Q $ as evaluated from the minimum saturated quantum value of the momentum using the Heisenberg principle. That is to say, eq.(\ref{Qmaxbecb}) is a {\it robust} result. 
In conclusion, a gravitationnally bounded BEC \textbf {cannot be the DM}.}
\end{itemize}

\section{Thermal and non-Thermal Axions}

The main DM candidate for a scalar particle forming a BEC condensate is the axion \cite{axion,tres}. For a report on axionic DM and axion like particles (`ALPs') see for example \cite{alps}.

In the usual scenario of DM axions, axions decouple soon after the QCD phase transition ($ g_d \sim 25 $) and
then they are assumed (i) to become nonrelativistic, (ii) to thermalize and (iii) to form a BEC \cite{sik}.
(Ref. \cite{sdav} recently criticized this scenario). Hence, the bound eq.(\ref{conBE}) clearly shows that
\textbf{no DM axion-like BEC can be formed.} 

{\vskip 0.1cm}

\textbf{For non-thermal axions}, the canonical scenario is the axion vacuum misalignment scenario \cite{kt},\cite{tres},\cite{mt},\cite{gorbu},\cite{alps}; see also refs \cite{co} - \cite{co-hall} for more recent discussions.  In this scenario, the axion field, denote it $\bar \theta$, is not initially at the minimum of its potential and is thus "misaligned" with it. When the axion mass is around a temperature of $T \sim \Lambda_{QCD}$, the axion field will roll toward the minimum, will overpass it and thereafter it will oscillate, these are like coherent oscillations.  The axion potential is negligibly small at  Temperatures $ T >> T_ {QCD}$,  the effective potential vanishing at high temperature. As $T$ decreases, the field starts to roll down  from its initial value towards its zero value at the minimum of the potential ($ \bar\theta = 0$) and the axion mass starts to be generated. 
The temperature dependence of the axion mass follows the relation
\be 
m (T) \; \simeq \; 0.1 \; m \;(T = 0)\;  \left( \;\frac{\Lambda_{QCD}}{T}\;\right)^{3.7}
\ee

The characteristic axion momentum at birth is $m (T_1)$. $T_1$ is the temperature when oscillations begin, defined as $m (T_1) = 3H (T_1)$. For an axion of mass $m \sim 10^{-5}$ eV, $T_1$ is about $1$ GeV, and in general $T_1$ scales as $ m^{+0.18}$.  The initial axion number density is given by
\be
n(T_1) \; \simeq \; \rho (T_1) / m(T_1)\;
\simeq \;m(T_1)\; \bar\theta_1^2\;(f/N)^2 /2
\ee
Without significant entropy production since the starting of axion oscillations, the axion's contribution to the energy density today can be derived from the ratio $( n/s ) $ of the initial axion number density to entropy density multiplied by the entropy density today $s_0$ times the axion mass $m$. (If significant entropy have been produced, the increasing entropy factor in a comoving volume will reduce the present axion mean density by the same factor). 
Taking into account anharmonic oscillation effects in the axion motion, the average axion's contribution to the  energy density today is given by 
\be \label{rhoa}
\rho_{DM} = \rho_{crit} \; 0.13 \times 10^{\pm 0.4} \; \Lambda_{200}^{-0.7} \; 
F({\bar \theta}_1) \; {\bar\theta}_1^2 \; \left(\frac{m}{10^{-5} \; \rm eV}\right)^{-1.18} \ee

{\vskip 0.3cm}

where $ \Lambda_{200} \equiv \Lambda_{QCD}/200 $ MeV, $\bar\theta_1 $ is the value of the axion phase field when 
axion oscillations begin. Its canonical value is $ \bar\theta_1 = \pi/\sqrt3 . \; F({\bar \theta}_1) $ accounts
for anharmonic effects, being $ F(\pi/\sqrt3) \simeq 1.3 $. 

{\vskip 0.2cm}

$ \Omega_{DM} = 0.22 $ in Eq.(\ref{rhoa}) yields for the axion mass:

\be \label{ma}
m = 3.77 \; 10^{-5} \; J^{-0.85} \; {\rm eV} \quad , \quad J \equiv  
\frac{3  \;\; {\bar \theta}_1^2 \;\; 10^{\pm 0.4} \; \;F({\bar \theta}_1)}{\pi^2 \; \Lambda_{200}^{-0.7} \; 1.3} \; \sim 1 \; .
\ee

{\vskip 0.1cm}

Axions are produced as semi-relativistic particles  at a Temperature much more larger than their mass. Axion production from the misalignement scenario is highly non-thermal and forms a zero-momentum population,  the phase occupancy is given by:
\be
f (p = 0)\; \simeq \; \frac{m \;(T_1) \; (f/N)^2}{ p_1 ^3} \; \sim \; 10^{50}\; \left(\;\frac{m}{eV} \;\right)^{-2.7}
\ee
 
These axion field oscillations truly correspond to a Bose  condensate and are like coherent oscillations.

Likewise, we can estimate the typical axion velocity today 
\be \label{vaxion}
v/c \;\sim \; (\;\frac{p}{m}\;)_{today} \; \sim \; \frac{p_1\;(3K/T_1)}{m}\;\sim\; 10^{-22}\; \left(\;\frac{m}{eV}\;\right)^{-0.82}
\ee

Moreover, we can also estimate the phase-space density $ Q $ using the
DM density  Eq.(\ref{rhoa}) and the axion velocity  Eq.(\ref{vaxion}) with the result

\be \label{meaq}
Q =  \rho_{crit} \; 2.86\times 10^{\pm 0.4}  \; \; 10^{66} \; \Lambda_{200}^{-0.7} \;
 \left(\;\frac{m}{\rm eV}\; \right)^{1.28}, 
\ee

which yields:
\be \label{meq}
\left(\frac{Q}{\rm keV^4}\right) = 3.21 \times 10^{49.4} \; \; 10^{\pm 0.8} \;
\left(\;\frac{m}{\rm eV}\;\right)^{1.28} \; 
\ee
\be \label{mem}
m = \; 4.8 \;\; 10^{-39} \; {\rm eV} \; \;  10^{\mp 0.6}\;\left(\frac{Q}{\rm keV^4} \right)^{0.78} \; 
\ee

{\vskip 0.1cm}

 From the well established different sets of  galaxy observations \cite{datos}, \cite{datos2}, \cite{datos3} and their compilation \cite{astro} Eq.(\ref{Qdato}), $Q_{\rm today} \leq Q$ is in the range: 
\be\label{Qdato2}
1,12 \times 10^{-8}\; < \;\left(\frac{Q_{\rm today}}{\rm keV^4}\right) \;< 1.65 \; ,
\ee
which implies the axion mass values:
\be\label{mam2}
m = \; 5 \;\times \; 10^{-45} \; \rm{eV} \;\; -  \;\; 7 \; \times \; 10^{-39} \;\rm eV
\ee
 
We see from Eqs.(\ref{mam2}) and (\ref{ma}) that the axion mass in the misalignement scenario is hugely larger than the required values to satisfy the galaxy phase density data.

\medskip

In addition, in order to reproduce the observed galaxy structures, from the non-linear regime computations eg. \cite{hubagru}, \cite{sik}, Eq.(\ref{22}), the axion mass needs to have typically the order of magnitude  $ m \sim 10^{-22}$ eV.
We see from Eqs.(\ref{22}) and (\ref{ma}) that the axion mass in the vacuum misalignment scenario is 
{\bf 17 orders of magnitude too large} to reproduce the observed galactic structures.

\medskip

In summary, the axion in the misalignment scenario cannot be the total dark matter, two robust constraints point to the same conclusion: On the one hand, the axion mass is too large by huge orders of magnitude to satisfy the observed phase space density galaxy data, and on the other hand the axion mass is too large too (by 17 orders of magnitude)  to describe the observed galaxy structures. 

\medskip

We considered in this section the axion in the canonical misalignment  scenario. More recent models can be considered by taking into account various effects as: (i) symmetry breaking with parametric resonance \cite{co}, \cite{harigaya}, (ii) anharmonicity effects for a initial value $\bar\theta_i$ approaching $\pi$ because of fine tuning or special inflationary dynamics \cite {gonza}, \cite{taka}, (iii) a "kinetic misalignment" mechanism \cite{co-hall}, in which the axion initial kinetic energy can be larger than the potential energy, thus delaying the onset of axion field oscillations. However, the fundamental picture of axions in the basic misalignment scenario as a coherent zero-momentum condensate of coherent oscillations does not change in all these models (i), (ii), (iii) nor the main range of their velocity and all these models produce axion masses in the range $(0.1 - 100)\; 10^{-3}$ eV. These variations do not change at all the robust axion results found here, as these results do not depend on the details of the particle models, and the exclusion constraints on the axion masses are on huge orders of magnitude. 

\section{Conclusions}

Present experimental limits leave as available window 
for the axion mass \cite{sik}
\be\label{mexpax}
 6 \times 10^{-6} \; {\rm eV}\; < \;m_a \;< \;2 \times 10^{-3}  \; {\rm eV} \;   
\ee

\begin{itemize}

\item{The window eq.(\ref{mexpax}) {\bf disagrees} by many orders of magnitude both with the galaxy phase-space density constraint
 Eq.(\ref{conBE}) and with the required value $ m \sim 10^{-22}$  galaxy structure constraint in order for the axion to be DM.}

\item{The existence of the axion particle is well motivated from QCD \cite{axion},\cite{tres}. 
But, as we have seen, the axion {\textbf cannot be the DM particle}. The two observables: the average DM density $ \rho_{DM} $ in real space and the phase space density $ Q $ 
robustly constrain in an inescapable way both: the possibility to form a BEC, eg $( T_d / T_c )$, and the DM particle mass $ m $  ruling out BEC DM in general,
and the BEC axion DM in particular.}
\item{Moreover, the value $ m \sim 10^{-22} $ eV can only be obtained with a number of ultrarelativistic degrees of freedom at
decoupling {\bf in the trillions} which is impossible for decoupling in the radiation dominated era.}

\item{In addition, we have also considered inhomogenous gravitationally bounded BEC's 
supported by the bosonic quantum pressure independently of any particular particle physics scenario. 
For a typical size $ R \sim $ kpc and compact object masses $ M  \sim 10^7 \; M_\odot $ they remarkably lead to 
the same particle mass  $ m \sim 10^{-22} $ eV  as the BEC free-streaming length. However, 
the {\bf phase-space density} for the gravitationally bounded BEC's turns out to be {\bf more than sixty orders of magnitude smaller} 
than the galaxy observed values.} 

\item{We have provided here a generic treatment, independent of the particle physics model and which applies to all DM BEC, in  both: in or out of equilibrium situations. We conclude that the BEC cannot 
be the total DM. The axion can be candidate to be only part of the DM of the universe.}

\item{In all the DM BEC discussion here it is assumed that axions represent the whole
DM in the universe, as is usually the case to investigate the feasibility of a DM candidate. In mixed scenarios where particles other than axions could form a large part of the DM, one could have an axion DM BEC constituing a part of the universe DM.} 

\item{In supersymmetric models the supersymmetric partner of the axion
is a fermion called axino, degenerate in mass with the axion.
An {\bf axino} with mass {\bf in the keV scale} would be a good warm dark matter (WDM)
candidate. Actually, an axion (and hence an axino) with particle mass in the keV scale
naturally appeared for a decoupling temperature $ T_d \sim 10^{11} $ GeV, [see eqs.(\ref{conBE})-(\ref{mTE})].}

\item{We would like to stress that although not being the DM,  the axion may play a crucial role in cosmology.
The observed dark energy density $ \rho_{\Lambda} = (2.35 \; {\rm meV})^4 $
indicates an energy scale in the meV $ = 10^{-3} $ eV. This energy value is in the allowed window of the axion masses.
Therefore, the axion may be the source of the dark energy through the zero point
cosmological quantum fluctuations as we derived in  Ref. \cite{de}.
In addition, white dwarf stars observations  would suggest axions in the range of $2\;-\; 8$ meV \cite{isern}, \cite{wang}.}

\item{Overall, a robust conclusion of this paper is that the BEC in general, and the BEC axion in particular, cannot be the total Dark Matter of the Universe. However, they can play an important role in astrophysics and cosmology. We see indications for an axion mass in the meV range from dwarf stars observations eg  \cite{isern},  \cite{wang},
and mainly from the dark energy scale as we studied in Ref \cite{de}.  In addition, the misalignment scenario \cite{kt}, \cite{tres}, \cite{mt} may be able to produce axions with mass in the meV range.} 
\end{itemize}

{      
\end{document}